# Protonation States of Remote Residues Affect Binding-Release Dynamics of the Ligand but not the Conformation of apo Ferric Binding Protein

Gokce Guven, Ali Rana Atilgan, Canan Atilgan*

Sabanci University, Faculty of Engineering and Natural Sciences,

Tuzla, 34956 Istanbul, Turkey

*Corresponding author

Faculty of Engineering and Natural Sciences
Sabanci University, Tuzla 34956, Istanbul, Turkey
*e-mail*: canan@sabanciuniv.edu
*telephone*: +90 (216) 483 9523
*telefax*: +90 (216) 483 9550

# ABSTRACT

We have studied the apo ($Fe^{3+}$ free) form of periplasmic ferric binding protein (FbpA) under different conditions and we have monitored the changes in the binding and release dynamics of $H_2PO_4^-$ that acts as a synergistic anion in the presence of $Fe^{3+}$. Our simulations predict a dissociation constant of 2.2±0.2 mM which is in remarkable agreement with the experimentally measured value of 2.3±0.3 mM under the same ionization strength and pH conditions. We apply perturbations relevant for changes in environmental conditions as (i) different values of ionization strength (IS), and (ii) protonation of a group of residues to mimic a different pH environment. Local perturbations are also studied by protonation or mutation of a site distal to the binding region that is known to mechanically manipulate the hinge-like motions of FbpA. We find that while the average conformation of the protein is intact in all simulations, the $H_2PO_4^-$ dynamics may be substantially altered by the changing conditions. In particular, the bound fraction which is 20% for the wild type system is increased to 50% with a D52A mutation/protonation and further to over 90% at the protonation conditions mimicking those at pH 5.5. The change in the dynamics is traced to the altered electrostatic distribution on the surface of the protein which in turn affects hydrogen bonding patterns at the active site. The observations are quantified by rigorous free energy calculations. Our results lend clues as to how the environment versus single residue perturbations may be utilized for regulation of binding modes in hFbpA systems in the absence of conformational changes.

# AUTHOR SUMMARY

Bacterial ferric binding protein A (FbpA) plays a major role in iron confiscation from its host. Phosphate has been long identified as a synergistic anion for iron in this transport protein, while its mode of action is still debated. We have demonstrated in this work that phosphate binding to apo FbpA is a heavily environment dependent activity. Hydrogen-bond network motifs between the phosphate and its surrounding residues in FbpA become more stable when we lower the pH. With the aid of extensive molecular dynamics simulations conducted for different protonation scenarios, we show that the average structure of the protein is not altered during phosphate binding. Despite the lack of conformational transitions, however, phosphate binding-release dynamics is substantially affected by these changes. We find that the charges are redistributed, giving rise to altered fluctuation patterns of the loops remotely controlling the binding activity. Our free-energy estimates compare extremely well with those of experimental studies. While 0.9 bound fraction is observed by tuning the pH to low values, compared to the value of 0.2 in the wild type, a single-residue mutation located on a distant surface loop, D52A, enhances the phosphate binding fraction to 0.5, for which the hydrogen bond network in the vicinity of the phosphate considerably loses its stability. The exact conditions in the periplasmic space of gram-negative bacteria are not well characterized, but are known to be much different than those in the cytosol. Our work offers a new way of perceiving the local environment as being capable of inducing substantial changes in functional dynamics of individual proteins while suppressing conformational transitions.

# INTRODUCTION

In ligand-binding related problems, the role of the flexibility of the receptor is gaining more attention with the advent of new sampling methodologies and hardware/software breakthroughs for advanced simulation techniques [1]. Although the importance of flexibility has been recognized, especially with the development of conformational selection model for binding [2,3], the traditional approach to binding problems utilizes docking and scoring functions[4]. More recently, accelerated molecular dynamics and metadynamics have been used to identify the lowest energy binding modes of the most potent ligand [5]. To predict binding affinities with precision, a correct description of the free energy of binding must be made. This is a relatively easier task once the bound conformation is correctly defined using a host of free energy estimation techniques [6]. Perhaps more interesting for direct applications to real life problems is a correct prediction of the $k_{on}$ values which requires multiple observations of the binding event starting from different solvated poses of the ligand [7]. In this context, it has been shown that conventional molecular dynamics (MD) simulations may be used to predict target binding sites of the drug dasatinib and kinase inhibitor PP1on Src kinase on the time scale of microseconds [8], albeit with the observation of single binding events during the course of the simulation. The question of quantitative prediction of binding/release kinetic parameters through MD simulations remains to the evaluation of systems whereby multiple events are observed within the accessible time scales.

Here, we treat the specific case of phosphate binding to bacterial ferric binding protein A (FbpA) which plays a central role in iron acquisition from the host in gram negative bacteria. For survival, these organisms compete for iron in the host either bound to transport proteins such as transferrin or tight binding chelators [9]. Phosphate is the native synergistic anion in recombinant neisseria FBP from *E. coli*, [10] and possibly also in other gram negative bacterial species [11], but its exact role in iron sequestration from the outer membrane proteins, transport across the periplasm and release on the cytosolic side is debated. The crystallographic structures of both the apo and holo forms of FbpA carry a phosphate group in the $Fe^{+3}$ binding site, albeit with shifted interaction partners [12,13]. The main mode of motion of FbpA is hinge bending of two domains [14]. The holo form is a closed conformation whereby residues from both domains arrange around the iron which establishes an octahedral geometry by utilizing phosphate and water as well as protein contacts. Apo-FbpA displays an open conformer whereby the phosphate is in contact with residues on only one domain.

An active role was attributed to phosphate in removing non-transferrin bound iron from the host, particularly in its efficiency towards iron from low-affinity chelators such as citrate [15]. While FbpA mutants defective in binding phosphate are nevertheless capable of binding iron so that growing in ferric citrate media of the bacterial strain is possible [16], phosphate is known to enhance the apparent binding affinity of $Fe^{+3}$ by one order of magnitude [17]. Furthermore, FbpA is able to carry iron in an open form in the absence of phosphate [18]; thus, the main role of the anion is to stabilize the closed conformation of FbpA, possibly by preorganizing the iron binding site [17].

The extremely high affinity of FbpA for iron with a $K_a$ value of $10^{18}$ $M^{-1}$ at the assumed pH of 6.5 of the periplasm [19] presents the problem of efficient release at the cytosolic membrane receptor. This task may also require an active role for the synergistic anion. The proton-assisted dissociation of the phosphate seems to be a trigger for subsequent dissociation of iron in a two-step process [20]. Furthermore, it was shown that anion exchange provides differential control over iron binding affinity of FbpA [21]. By utilizing the anionic alternatives present in the rich periplasmic space, gram negative bacteria may thus modulate the delivery of iron to the cytoplasm on demand. In strict contrast to holo-FbpA, phosphate is only weakly bound to apo-FbpA with a $K_d$ value of 2.3±0.3 mM at the assumed pH of 6.5 of the periplasm [12]. However, a linear correlation between anion dissociation constants in apo-FbpA and anion exchange rates in holo-FbpA prevailing over three orders of magnitude suggests similar molecular level processes for both events [22].

In this study, we investigate in detail the direct interactions between the protein and the anion in the absence of iron using extensive molecular dynamics (MD) simulations. Since the role of phosphate as a synergistic anion is much debated, we pursue the scenario that the rich environmental conditions of the periplasmic space, with its wide distribution of ionic species maintaining a 30 mV Donnan potential [23] and its gel-like properties harboring microenvironments reserved for specific tasks [24], may be utilized to optimize the binding kinetics of FbpA. Moreover, the periplasmic pH (unlike that of the cytoplasm) shifts according to external pH, with no apparent homeostasis for neutralophilic bacteria [25] offering periplasmic proteins subtle control of binding-release kinetics of anions by utilizing the changes in the external environment. To investigate the consequences of the routes that may have evolved in the periplasm for control of binding/release of ions, we explore alternative circumstances under which phosphate dynamics are modified by perturbing (through protonation or mutation), remote residues which have the propensity to be protonated. These include the four histidine residues of FbpA and D52 which has an upshifted pKa value according to PHEMTO server calculations ([26], and figure 6 in [27]). D52 has also been implicated as an efficient remote coordinator of the binding site located ca. 30 Å away [28]. We delineate the differences in the dynamics of the protein, in comparison to the wild type, caused by these manipulations.

Our aim in this manuscript is two-fold: On the one hand, we show that the binding-release kinetics of a small, rigid ligand may be correctly modeled by MD simulations on the time scales (200 ns) currently accessible by moderate computer power. This expands on the possibility of modeling binding events and predicting affinity using conventional MD simulations. On the other hand, for the specific case of FbpA, we show that phosphate affinity may be tuned by moderating charge states of distal sites even in the absence of $Fe^{+3}$ ion. We further evaluate the molecular level processes that lead to the altered dynamics.

## METHODS

**Haemophilus influenza periplasmic ferric binding protein systems (FbpA).** FbpA consists of 309 amino acids made of a N-lobe (residues 1-82, 88-101, 226-276) and C-lobe (residues 83-87, 102-225, 277-309). Ferric ion binds between the two lobes via a hinge motion, and is coordinated by a synergistic phosphate anion in the holo x-ray structure (Protein Data Bank (PDB) code 1MRP) as well as the active site residues of Y195, Y196, E57, H9 and a water molecule [13]. The apo form x-ray structure (PDB code 1D9V) [12] has the phosphate anion only, and is coordinated by S139, K140, A141, N175, Y195 and, Y196 (figure 1).

**System Construction for Molecular Dynamics Simulations.** MD simulations on apo FbpA are carried out under a variety of conditions. These include physiological ionic strength of 150 mM having 25 $Na^+$ and 25 $Cl^-$ ions, versus those systems that are neutralized by only three $Na^+$ ions. Wild type FbpA runs are carried out with the ionizable groups set to their protonated state according to their standard $pK_a$ values; these runs are labeled WT. We assume that these conditions are similar to that of the periplasm in terms of pH set to 6.5. This selection is made because the active transport of metals in gram negative bacteria is coupled to the electrochemical proton gradient and the hydrolysis of ATP which causes the periplasm to be slightly acidic [29]. In other runs, local changes are introduced. These may be in the form of the D52A point mutation, as well as protonation of the same residue, $D52^+$, which has an upshifted pKa value and has been implicated in the mechanical manipulation of the hinge motion in previous computational analysis as mentioned in the Introduction. We also account for the protonation propensity of histidine residues and in another set of simulations all four histidine residues in the protein along with D52 have been protonated (at the $N_\delta$ position for histidine and carboxyl group in D52). These runs are labelled as $H^+D52^+$, and the locations of the perturbed residues are displayed in figure 1c. A list of all MD simulations performed in this work is provided in Table 1. Under the slightly acidic conditions of the periplasm, we model the phosphate group as $H_2PO_4^-$, since the pKa of $H_2PO_4^-$ to $HPO_4^{-2}$ shift is 7.2 [30]. The protein-water-phosphate systems are prepared using the VMD 1.8.6 program, autoionize and solvate plugins [31] and the NAMD software package is used for MD simulations [32]. The systems were neutralized with $Cl^-$ ions as necessary. All systems have at least two runs of 200 ns length each.

After soaking the protein in a water box such that there is at least 10 Å layer of water molecules in each direction from any atom of the protein to the edges of the box, the systems are neutralized with $Na^+$ and $Cl^-$ ions. The simulated protein-water complexes have approximately 9370-9640 water molecules. The CharmM22 force field parameters are used for protein and water molecules [33]. TIP3P model is used for water [34]. The binding parameters for the synergistic anion $H_2PO_4^-$ are taken from the literature [35].

Long range electrostatic interactions were calculated using particle mesh Ewald (PME) method [36]. The cutoff distance for non-bonded van der Waals interactions was set to 12 Å with a switching function cutoff of 10 Å. Rattle algorithm was used to fix the bond lengths to their average values. Periodic boundary conditions were used and the equations of motion were integrated using the Verlet algorithm with a step size of 2 fs [37]. Temperature control was carried out by Langevin dynamics with a dampening coefficient of 5/ps and pressure control was attained by a Langevin piston. Volumetric fluctuations were preset to be isotropic.

All systems were first subjected to energy minimization with the conjugate gradients algorithm until the gradient tolerance was less than $10^{-2}$ kcal/mol/Å. The final structures were then run in the NPT ensemble at 1 atm and 310 K until volumetric fluctuations were stable to maintain the desired average pressure. Finally, the runs in the NPT ensemble were extended for data collection. The coordinate sets were saved at 2 ps intervals for subsequent analysis.

**Free energy calculations.** We have carried out free energy of perturbation (FEP) calculations to estimate the binding free energy of phosphate. We follow the protocol outlined in [38] and express the salvation free energy as the sum of three terms: Annihilation of $H_2PO_4^-$ in the solvent, annihilation of $H_2PO_4^-$ in the bound conformation, and the free energy change of restraining to account for the loss of configurational entropy upon binding (see figure 8 in reference [38]). We choose two configurations from two different simulations both for the bound and solvated $H_2PO_4^-$ initial configurations. Both the forward (annihilation; $\lambda$: 1→0) and reverse (creation; $\lambda$: 0→1) calculations are carried out for each system. Separate decoupling of electrostatic and van der Waals interactions is utilized. While the latter are modified linearly with $\lambda$ throughout the simulation, the former are turned-off linearly in the range at $\lambda = [0, 0.5]$. The annihilation/creation process is divided into 32 increments ($\Delta\lambda = 0.03125$). At each increment, 50 ps equilibration is followed by 150 ps of data collection. The error bars are estimated using both the Bennett acceptance ratio [39] and simple overlap sampling methods [40]. Both approaches give the same error bounds in our calculations.

In the bound conformations, positional restraints are imposed on the $H_2PO_4^-$ to maintain the appropriate orientation in the binding site using collective variables module implemented in NAMD [41]. The atoms involved in the restraining are selected by monitoring the parts of the MD trajectory where the $H_2PO_4^-$ remains bound. We find that the hydrogen bonds between the two O atoms of $H_2PO_4^-$ and the H atoms on S139 (HG1), G140 (HN), A141 (HN), N175 (HD21/HD22), N193 (HD21/HD22), Y195 (HH) and Y196 (HH) provide a good descriptor with average center of mass distance between the two groups in the free MD simulations being 1.8±0.6 Å. We fix the collective variables restraining distance at 2.0±0.5 Å with the constraining force constant of 100 kcal/mol/Å$^2$.

The center of mass constraining potential is calculated by the from the Sackur-Tetrode equation, $\Delta G_{trans} = -k_B T \ln(c^0 \Delta V)$ where $c^0$ is the standard concentration of 1 L/mol, $\Delta V$ is the effective volume sampled by $H_2PO_4^-$ ($4\pi R^3/3$), with $R$ being the average collective variable distance recorded in the bound form in the unperturbed state. The rotational contribution, $\Delta G_{rot}$, is estimated by recording the positions of a selected O atom of $H_2PO_4^-$ (any of the four atoms works since it is a rigid molecule) after best fitting the trajectory on the same set of protein atoms used for restraining (see above) and removing the translations of the P atom. This process maps the O atoms on the surface of a sphere of radius equal to the P–O distance (for the solvated system the whole surface of the sphere is sampled). Given that the fraction of the covered surface is $f$, $\Delta G_{rot} = -k_B T \ln(f)$.

## RESULTS

**Phosphate binding dynamics is modulated by the system conditions.** In all the MD simulations, equilibration is readily achieved (see Table S1 for average root mean square deviation (RMSD) values of N- and C-lobe positions separately as well as the overall structure). Throughout the simulations, $H_2PO_4^-$ displays multiple binding and release events rendering it possible to estimate kinetic and thermodynamic constants. The average structures under the different conditions, i.e. pH variation, ionic strength variation, mutation/protonation of a given site, presence/absence of $H_2PO_4^-$, are the same within error bounds. We find that the RMSD of the N-lobe is slightly lower than that of the C-lobe (1.6±0.4 Å vs. 1.9±0.5 Å). The overall RMSD after equilibration is 2.3±0.4 Å for the total of over 2 μs simulations.

While the average structures do not vary, the binding and release dynamics differs markedly depending on the conditions. The bound fraction of $H_2PO_4^-$ for each simulation is listed in Table 1. We display the distance between the P atom of the $H_2PO_4^-$ group and N atom of amine group of N175 for three separate systems each with two independent 200 ns long runs in figure 2. We take phosphate to be bound if it is within the binding cavity (using a cutoff distance of 10 Å) and unbound otherwise. We find that for the wild type, the bound fraction is 20.1 % and 18.2 % in the two independent simulations. When the ionization states of the four histidines and D52 are modified to mimic a possible modification of the environment to the FbpA surface, the ion remains bound all the times in one case, and it is bound 87% of the time in the second case (we have prolonged the latter simulation for another 100 ns, and we find that the ion enters the binding cavity within the next 2 ns and is fully engaged within 10 ns.) Finally, modification of the ionization state of only D52, either as protonation or as alanine mutation, leads to a moderate bound fraction with the average value of 54.2±1.9% (it is 56.2 and 52.5% for the D52A simulations and 53.9 for the D52$^+$).

Treating the binding of phosphate to FbpA as a bimolecular reaction,

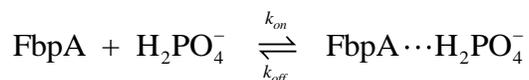

$$\text{FbpA} + \text{H}_2\text{PO}_4^- \underset{k_{off}}{\overset{k_{on}}{\rightleftharpoons}} \text{FbpA}\cdots\text{H}_2\text{PO}_4^-$$

we can determine the equilibrium constant for the dissociation of $H_2PO_4^-$ $K_d = k_{off}/k_{on} = [\text{FbpA}^-][H_2PO_4^-]/[\text{FbpA}\cdots H_2PO_4^-]$ from the fraction of trajectories where the ligand is bound/unbound. In the current case, we have 1:1 ratio of [FbpA$^-$]:[H$_2$PO$_4^-$]. We may also predictthe off-rate from the amount of time H$_2$PO$_4^-$ remains bound before dissociating completely. The precision of this quantity is lower since the number of observed release – rebinding events is limited. The thermodynamic and kinetic parameters that were calculated from the simulations are listed in Table 2.

Experimentally, dissociation constant measurements of phosphate from apo-FbpA were made under very low dilution conditions (15 μM) and a $K_d$ value of 2.3±0.3 mM was reported at a pH of 6.5 [12]. Our corresponding system has a $K_d$ of 2.17±0.19 mM, in excellent agreement with these UV difference spectrophotometry measurements. We also calculate the $K_d$ value for the same pH, yet at the physiological ionic strength of 150 mM in which case the trajectories converge to the equilibrium distribution on a longer time scale than that observed at 200 ns for the low ionic strength. We have three separate runs totaling of 600 ns, and their average $K_d$ = 2.7±1.6 mM, which is on the same scale as the experimental value, but with a much larger error bar. The calculated $K_d$ values for the D52 and H$^+$D52$^+$ perturbation are also listed in Table 2 and they reflect the longer time the ligand spends bound to the protein.

Free energy of binding for each system were estimated from the bound/unbound fractions via $\Delta G^{observed} = k_B T \ln K_D$. The FEP calculations made for the two extreme cases of WT and H$^+$D52$^+$ provide $\Delta G^{FEP}$ in excellent agreement with $\Delta G^{observed}$ values (Table 2). These values not only corroborate the observed binding fractions, but also lend clues on the factors that contribute to the shift in the equilibrium (Table 2). We find that the change in the protonation states of five residues do not affect the solvation energy of $H_2PO_4^-$, as expected. However, they are markedly different for binding in the cavity. The annihilation energy in the bound form is stronger for the $H^+D52^+$ system by 3.6 kcal/mol. This difference is somewhat offset by the loss of configurational entropy due to restraining imposed in the cavity. Both the rotational freedom and center of mass motions are more restricted for the $H^+D52^+$ system, leading to the final free energy difference of 1.8 kcal/mol. In the next subsection, we investigate the phosphate-protein interactions in the bound form so as to elaborate on the molecular details of ligand binding in this system.

**Local perturbations lead to changes in phosphate coordination due to significant electrostatics redistribution.** While the average conformation of the protein is the same under all conditions (Table S1), it is still possible that there are some significant local conformational changes, especially within the binding cavity, not reflected in the RMSD values. We therefore measure the solvent accessible surface area (SASA) of residues 139, 140, 141, 175, 193, 195, 196 which directly contact $H_2PO_4^-$ when fully engaged in the cavity. However, the accessibilities of these critical residues are also unchanged within error bounds; leading to an overall average and standard deviation of 253 ± 46 Å$^2$ (individual values for the separate systems are listed in Table 1). Note that the averages are reported over the entire trajectory, not just the bound subset. On the other hand, configurational entropy calculations reported in the previous subsection (rotational and translational contributions to $\Delta G^{FEP}$) implicate that the strength of the contacts established between the protein and the ligand must be modified under the different conditions studied.

We monitor single snapshots of $H_2PO_4^-$ in the bound state from the various trajectories. We find that the coordination of the guest by the protein at this local scale is markedly dependent on the simulation condition (figure 3a). This observation is directly quantified in the hydrogen bond occupancies obtained from the parts of the trajectories where the $H_2PO_4^-$ remains bound to the protein (Table 3). In both the WT and D52A systems, the interactions are centered on N175 and Y196, while S139 and G140 are additionally involved throughout the $H^+D52^+$ trajectories. The involvement of the latter affects the motion of the $H_2PO_4^-$ in the cavity with the center of mass motion reduced by 60% and the rotational freedom reduced by 40 %; these values are reflected in the $\Delta G^{FEP}$ calculations. Similar observations for phosphate binding were made in another bacterial protein family [42] where it was found that the phosphate ion is tightly bound to the protein via 12 hydrogen bonds between phosphate oxygen atoms and OH and NH groups of the protein. The authors calculated a positive electrostatic potential and concluded that the hydrogen bond network is probably the key that explains the very high affinity of binding.

We emphasize that this tightened binding effect observed at the local binding region is a result of the perturbations that occur far from the binding site. For the $H^+D52^+$ system, the closest protonated residue is H9, located 12 Å from the bound $H_2PO_4^-$. More interestingly, D52 is 30 Å away from the guest, and yet its protonation suffices to significantly shift the equilibrium.

To investigate how such remote modulation of binding/release dynamics is achieved in FbpA, we monitor the electrostatic distribution around the protein using the Adaptive Poisson-Boltzmann Solver (APBS) package [43]. In Figure 3b, sample electrostatics surfaces are displayed for the three systems studied. Note that these are snapshots taken at the end of 200 ns simulations, although different time points from the trajectories give surfaces with essentially identical properties. The main effect of lowering the pH is to enlarge the positive surface around the entrance region of the phosphate. More subtle, but significant, is the growth of the positive field around the cavity as a result of the neutralization of the single site D52.

The kinetic parameters estimated from the mean first passage times of the anion (Table 2) reveals a significant effect of the changed electrostatic distribution around the binding pocket. The enlargement of the positive electrostatic surface affects the off rates (which vary by up to two orders of magnitude). Hence, the altered charge distribution near the binding pocket affects the affinity between the bound anion and the receptor.

**Changes in the electrostatic surface distributions affect molecular mechanics at functional sites.** While the average structure remains the same in all of the systems studied in this work (Table S1), the deviations from the mean display significant differences (Figure 4a). In fact, it is possible to distinguish between non-specific sites that function as mechanical centers and those that take on more specific functional roles. For example, large fluctuations in the loop spanning residues 84 to 92 on the N-domain are ever-present, irrespective of the environmental conditions. These regions interact with the C-terminal loop spanning residues 281-290. The two loops, which also display significant cross-correlations (Figure 4b), comprise the hinge region around which the main motion of FbpA occurs. Their role is mainly mechanical, and there are no conserved residues in either loop. Also maintaining their relative fluctuations, albeit with lower values, are loops 202 – 208 and 235 – 240. Despite their relatively close residue indices, these two loops reside on the opposite side of the two domains. Having no conserved residues, they may be considered to be a part of the same machinery of the hinge bending motion of FbpA.

The largest changes in the fluctuations occur in the turn region comprised of residues 44 to 49. Interestingly, residues E45, G46 and T49 within this region are highly conserved although, to the best of our knowledge, no functional role has previously been attributed to them. For the WT system, these residues display the largest fluctuations while their motions are substantially damped after protonating five residues. D52 protonation/mutation gives an intermediate result. A similar observation is also made for loop residues 217 – 222 where the highly conserved G222 resides. Moreover, while both of these regions are highly correlated with all residues across the protein for WT, they remain correlated only with each other and the N-terminal hinge loop at the lower pH (Figure 4b).

For both the auto- and the cross-correlations, D52A simulations stipulate an intermediate scenario. Although the inter-residue interactions are affected by the altered protonation states of residues, we find no specific sequence of events that correlate the protonated site (D52) and the binding pocket. As in our previous study on $Ca^{+2}$ loaded calmodulin [44], the absence of an apparent communication pathway between the active site and the allosteric site supports the ensemble view of allostery [45].

**DISCUSSION**

FbpA provides an interesting case whereby, in the absence of conformational change, functional dynamics is manipulated by altering the electrostatic distribution (figure 3b) which in turn induces subtle differences in the residue fluctuations around their identical average positions (figures 3a and 4). We had previously encountered a similar example in chymotrypsin inhibitor 2 in unbound form in comparison to its complex with the large substrate subtilisin novo [46]. Although the conformation of the molecule in the two states is almost identical, with 0.6 Å

RMSD, residue fluctuations in the protein are greatly altered due to the effective propagation of perturbation and the presence of remotely controlling residues.

It is interesting to note that although the 309 residue FbpA contains 40 residues which are very highly conserved (figure 4a), 36 are located in constrained, low-fluctuating regions, while only four are in two flexible loops. The relative fluctuations of these two loops are modulated by the changed conditions, while none of the other flexible regions are affected. Thus, mechanical control is achieved by modulating the fluctuations of a few flexible regions that seem to harbor a set of conserved residues with no other apparent function. It is plausible that certain residues in FbpA have evolved to occupy positions in electrostatically susceptible and mechanically effective positions. We note that our perturbation-response scanning method [14,28] can pinpoint such positions, as exemplified here by D52 and another charged surface residue, E31 in our previous work on calmodulin [47]. In contrast to the current work, the latter study demonstrated that large conformational changes related to calmodulin function were induced by the perturbation of the single residue [44], or the protonation of a group of residues mimicking the low pH environment [48].

Protonation and pK changes in protein-ligand binding are related to phenomena such as pH dependence of binding affinity. Substantial pK changes far from the binding site appear to be a general feature of the binding process; nevertheless, many proteins appear to operate at a pH-optimum of ligand binding where proton uptake/release is avoided [49]. Currently, there is no detailed model to identify how modification of the protonation states of a subset of residues in FbpA may be achieved under physiological conditions. However, the properties of the periplasmic space, whose composition is distinct from that of the surrounding media, provide an excellent environment for such alterations. The periplasm contains a wide distribution of ionic species so that it maintains a Donnan potential of ca. 30 mV, negative inside [23]. The motility of the dissolved substances in the gel-like periplasmic space is drastically reduced to about 0.1% of their values in aqueous solution, and it may well contain microenvironments serving specific tasks [24]. Moreover, the periplasmic pH (unlike that of the cytoplasm) shifts according to external pH, with no apparent homeostasis for neutralophilic bacteria [25]. Consequently, fluctuations in the external environment may also translate into subtle control of binding-release kinetics by periplasmic proteins.

This study also demonstrates that accounting for the flexibility of the protein, even in the presence of a completely rigid ligand and the absence of apparent conformational change in the receptor, is necessary to have a complete picture of ligand binding and release events [3]. In particular, remote control of the binding region dynamics by the environmental conditions via non-bonded interactions must be considered. The strength of conventional MD simulations and free energy calculations is evident; yet, methods that promote efficient conformational search of the ligand binding site need be developed [1]. It has been observed that when drugs are bound in their native pose, the fluctuations in the ligand conformations are remarkably more constrained than when they associate with alternative sites on the receptor [8]. It was proposed that alternative target binding pockets may thus be determined via MD simulations. Our current results might lead to a similar use of flexibility (or lack thereof) within the binding pocket to determine the conditions best describing the environment that optimizes the functioning of the receptor.

**Acknowledgments.** We thank Deniz Sezer for valuable discussions. This work was supported by the Scientific and Technological Research Council of Turkey Projects (grant numbers 110T624 and 113Z408).

**TABLES**

**Table 1. Summary of system parameters and perturbation scenarios**

| System label | Perturbation Scenario | Ionic strength (mM) | Number of atoms | Equilibrated Box size (Å) | Simulation length (ns) | Bound fraction, $p_{bound}$ | SASA of binding residues (Å$^2$)[b] |
|---|---|---|---|---|---|---|---|
| WT | -- | 0 | 32963 | 82×84×90 | 200 | 0.201 | 261 (50) |
| | | | | | 200 | 0.182 | |
| | | | | | 100[a] | - | |
| H$^+$D52$^+$ | H9, D52, H109, H218, H298 are neutralized | 15 | 33060 | 81×69×58 | 200 | 1 | 249 (43) |
| | | | | | 200[c] | 0.871 | |
| D52A | Mutation of a remote site | 3 | 33067 | 80×70×58 | 200 | 0.562 | 252 (44) |
| | | | | | 200 | 0.525 | |
| D52$^+$ | Protonation of a remote site | 3 | 33067 | 80×70×58 | 200 | 0.539 | 238 (48) |
| WT$_{150mM}$ | Increasing ionic strength | 133 | 33835 | 80×70×58 | 200 | 0.088 | 248 (45) |
| | | | | | 200 | 0.402 | |
| | | | | | 200 | 0.190 | |
| | | | | | 60[a] | - | |
| D52$^+_{150mM}$ | Mutation of a remote site, increasing ionic strength | 138 | 32963 | 80×70×58 | 200 | 0.159 | 268 (47) |

[a] Control runs where the phosphate group is initially solvated and are monitored for binding the cavity. Since these runs are halted after binding is observed (i.e. a single event), bound fraction is not reported.

[b] Calculated using SASA plugin of VMD on residues 139, 140, 141, 175, 193, 195, 196 over all the trajectory, irrespective of phosphate being bound or dissociated; standard deviation reported in parentheses

[c] At the end of 200 ns phosphate is unbound in this run (figure 2); it has been continued and rebinding is confirmed to occur within 10 ns.

**Table 2. Thermodynamic and kinetic parameters** [a]

| System label[b] | $K_d$ (mM) | $\Delta G^{observed}$ (kcal/mol) | $\Delta G^{FEP}$ (kcal/mol) | | | | | $k_{off}$ (s$^{-1}$) [b] |
|---|---|---|---|---|---|---|---|---|
| | | | Bound (creation) | Solvated (annihilation) | Translational | Rotational | Total | |
| WT | 2.17 (0.19) | **0.48** | -93.25 (0.22) | 89.16 (0.14) | 3.03 (0.02) | 1.19 (0.11) | **0.13 (0.49)** | 1.6×10$^8$ (0.4) |
| H$^+$D52$^+$ | 0.07 | **-1.64** | -96.89 (0.23) | 89.13 (0.12) | 4.45 (0.03) | 1.63 (0.13) | **-1.68 (0.51)** | < 5×10$^6$ |
| D52A, D52$^+$ | 0.43 (0.02) | -0.52 | | | | | | 2.7×10$^8$ (0.9) |
| WT$_{150mM}$ | 2.7 (1.6) | 0.61 | | | | | | 1.3×10$^8$ (0.7) |
| D52$^+_{150mM}$ | 2.7 | 0.61 | | | | | | 1.5×10$^8$ (0.4) |

[a] Uncertainty in reported values is provided in parentheses
[b] Calculated from the inverse mean first passage times

**Table 3. Per cent occupancy of hydrogen bonds between H$_2$PO$_4^-$ (acceptor) and protein side-chain atoms (donor)**

| Donor | WT | H$^+$D52$^+$ | D52A |
|---|---|---|---|
| SER139 | 15 | 30 | 14 |
| GLY140 | 2 | 29 | 7 |
| ALA141 | 12 | 5 | 6 |
| ASN175 | 27 | 24 | 29 |
| ASN193 | - | 3 | 1 |
| TYR195 | - | 3 | 3 |
| TYR196 | 17 | 13 | 20 |

**FIGURE LEGENDS**

**Figure 1. (a)** Apo hFbpA (PDB ID:1D9V) displaying residues that coordinate $H_2PO_4^-$ in the x-ray structure. N-terminal domain (blue), C-terminal domain (gray), $H_2PO_4^-$ (space filling) and its coordinating residues (stick) are displayed). Perturbed residues D52 (yellow) and the four histidines (purple) which were neutralized in the $H^+D52^+$ runs are also shown as sticks. **(b)** Close-up view of residues coordinating $H_2PO_4^-$. N193 is not displayed for clarity.

**Figure 2.** Distance between the P atom of $H_2PO_4^-$ and N (amino group) atom of residue N175 during the simulations. Two runs of 200 ns length is displayed for the WT (left), $H^+D52^+$ (center), and D52A (right) systems. $H_2PO_4^-$ is considered bound in the cavity when the distance is less than 10 Å (marked by the dashed lines). For the WT system, $H_2PO_4^-$ is bound ca. 20% of the time in both cases. For the system, $H_2PO_4^-$ remains bound for the whole duration of the simulation in run 1, and it is bound nearly 90% of the time in run 2. Note that we have continued run 2 beyond the 200 ns displayed and observed rebinding within 10 ns. The D52A (as well as the $D52^+$ system not displayed here) system displays an intermediate scenario, where $H_2PO_4^-$ remains bound ca. 50% of the time in both runs with many dissociation – association events observed through.

**Figure 3. (a)** Superposed snapshots of the motion of bound $H_2PO_4^-$ under different conditions. 25 consecutive snapshots spanning a time range of 5 ns are displayed in each case. The orientation is the same as figure 1b. Residues S139 (yellow), G140 (white), A141 (white), N175 (olive), N193 (olive), Y195 (green) and Y196 (green) are shown in stick representation while $H_2PO_4^-$ is shown in space filling. When bound, $H_2PO_4^-$ is gripped tighter by the surrounding residues in the perturbed systems compared to the WT. **(b)** Snapshots of electrostatics isocontours of FbpA drawns at ±0.5 $k_BT/e$. Blue is positive and red is negative. The orientation of the protein the same as in figure 1a. The positively charged field around the binding cavity grows considerably in the $H^+D52^+$ scenario. For D52A, this region somewhat grows but the main effect comes from residue 52 switching from a negative to a positive isocontour upon neutralization.

**Figure 4.** (a) Root mean square fluctuations averaged over 20 equally spaced trajectory pieces of 20 ns each. Highly conserved residues (according to their CONSURF [50] score of 9) are shown in red/green: those residing in secondary structural elements are in red (D52 is the larger red dot), and those in flexible loops are in green. Despite the average structure remaining the same in all the simulations, fluctuation patterns are markedly different under different conditions. A cartoon of the secondary structural elements is displayed on the top (α helices in red, β sheets in green, other secondary structurl elements in blue). (b) Residue cross-correlations. The diagonal of these figures constitute the RMSF figures in part (a). The correlations between different regions also change considerably, the cross-talk between different parts being highly reduced as more residues are protonated.

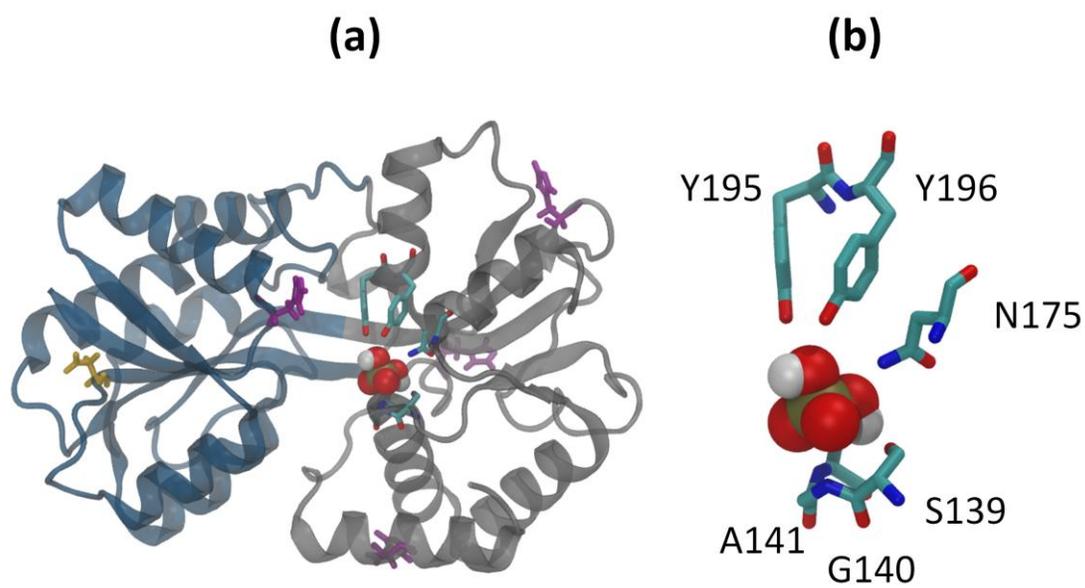

**Figure 1**

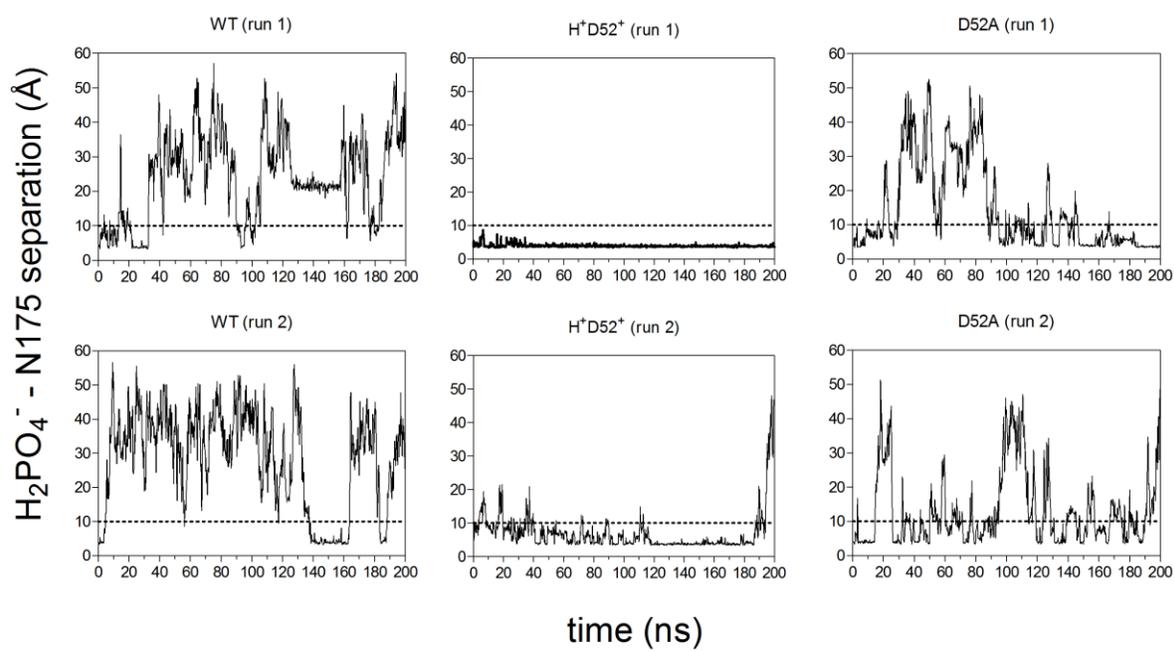

**Figure 2**

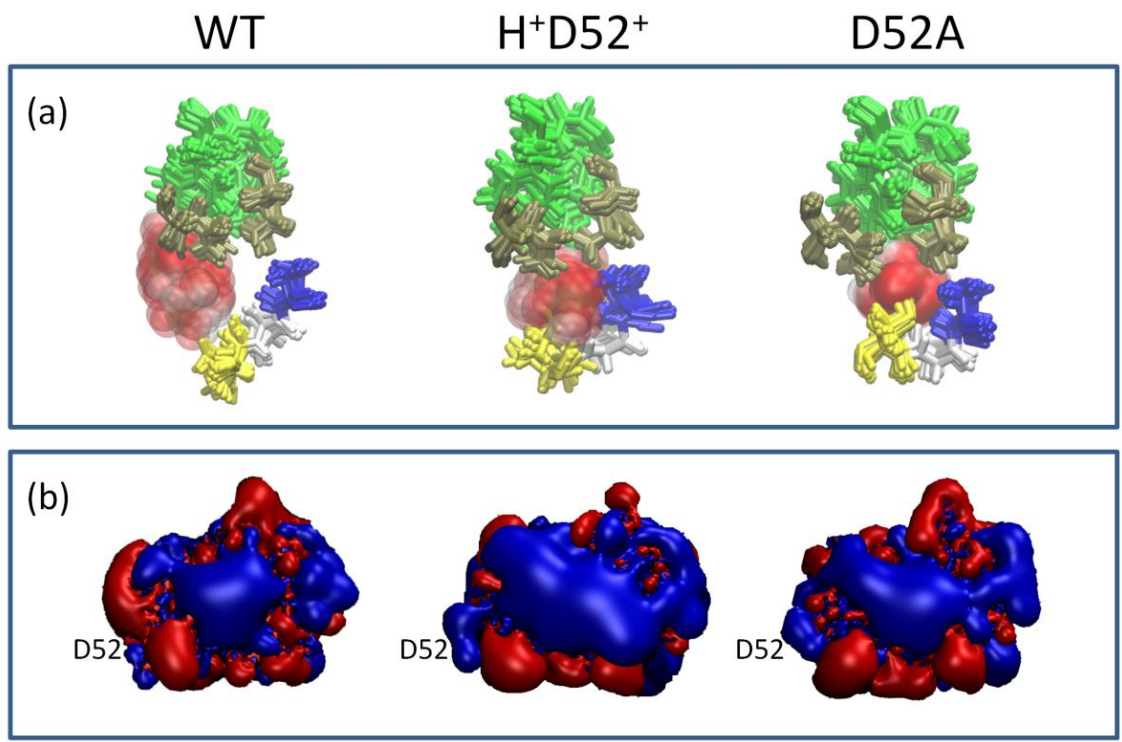

Figure 3

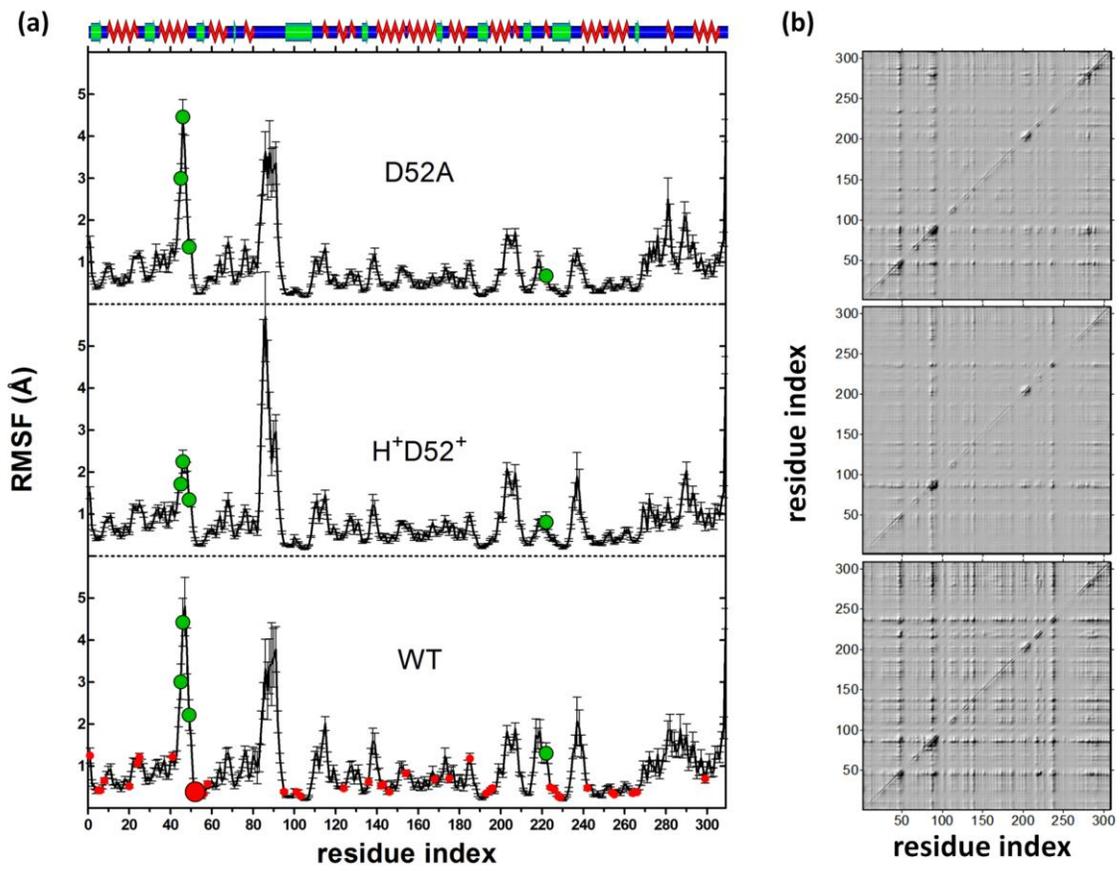

Figure 4

**Table S1.** Average backbone RMSD from the x-ray structure achieved during MD simulations listed in Table 1.

| System label | RMSD (Å) (equil. state) | RMSD (Å) N lobe | RMSD (Å) C lobe | Simulation length (ns) |
|---|---|---|---|---|
| WT | 2.4±0.4 | 1.3±0.1 | 1.8±0.4 | 200 |
|  | 2.1±0.3 | 1.7±0.3 | 1.8±0.3 | 200 |
|  | 1.9±0.4 | 1.2±0.2 | 1.5±0.3 | 100 |
| $H^+D52^+$ | 1.9±0.4 | 1.2±0.1 | 1.6±0.3 | 200 |
|  | 2.1±0.4 | 1.4±0.4 | 1.7±0.3 | 200 |
| D52A | 2.0±0.4 | 1.3±0.2 | 1.5±0.4 | 200 |
|  | 2.1±0.3 | 1.4±0.2 | 1.6±0.3 | 200 |
| $D52^+$ | 2.5±0.4 | 1.3±0.2 | 1.5±0.3 | 200 |
| $WT_{150mM}$ | 2.6±0.7 | 1.7±0.4 | 2.6±0.7 | 200 |
|  | 2.0±0.3 | 1.6±0.3 | 1.3±0.1 | 200 |
|  | 2.0±0.4 | 1.4±0.3 | 1.5±0.2 | 200 |
|  | 1.7±0.3 | 1.0±0.2 | 1.6±0.3 | 60 |
| $D52^+_{150mM}$ | 2.6±0.4 | 2.2±0.3 | 2.6±0.5 | 200 |